\documentclass[osajnl,amsmath,amssymb,twocolumn,superscriptaddress]{revtex4-1}%
\usepackage{amsmath}
\usepackage{xcolor}
\usepackage{graphicx}% Include figure files
\usepackage{dcolumn}% Align table columns on decimal point
\usepackage{bm}% bold math
\usepackage{float}
\usepackage{epsfig}% FOR CORRECT FIGURE CAPTIONS
\usepackage{amsmath}%
\usepackage{amsfonts}%
\usepackage{amssymb}
\usepackage{natbib}
\usepackage{soul}
\usepackage{mhchem}
\usepackage{color}

\usepackage{amssymb}
\usepackage{scalerel}
\usepackage{appendix}
\usepackage{lettrine}
\usepackage{lineno}

\begin{document}
\title{Photonic Snake States in Two-Dimensional Frequency Combs}
% Frequency Comb Replicas in Cylindrical Micro Resoffnators
\author{Salim B. Ivars}
%\email{salim.benadouda@upc.edu}
\affiliation{Institut Universitari de Matem\`{a}tica Pura i Aplicada, Universitat Polit\`{e}cnica de Val\`{e}ncia, 46022 Val\`{e}ncia, Spain}
%\affiliation{ICFO--Institut de Ciències Fotòniques, The Barcelona Institute of Science and Technology, 08860 Castelldefels (Barcelona), Spain}
\affiliation{Departament de F\'{i}sica, Universitat Polit\`{e}cnica de Catalunya, 08222 Terrassa (Barcelona), Spain}
\author{Yaroslav V. Kartashov}
\affiliation{ICFO--Institut de Ciències Fotòniques, The Barcelona Institute of Science and Technology, 08860 Castelldefels (Barcelona), Spain}
\affiliation{Institute of Spectroscopy, Russian Academy of Sciences, Troitsk, Moscow, 108840, Russia}
\author{P. Fern\'andez de C\'ordoba}
\affiliation{Institut Universitari de Matem\`{a}tica Pura i Aplicada, Universitat Polit\`{e}cnica de Val\`{e}ncia, 46022 Val\`{e}ncia, Spain}
\author{J. Alberto Conejero}
\affiliation{Institut Universitari de Matem\`{a}tica Pura i Aplicada, Universitat Polit\`{e}cnica de Val\`{e}ncia, 46022 Val\`{e}ncia, Spain}
\author{Lluis Torner}
\affiliation{ICFO--Institut de Ciències Fotòniques, The Barcelona Institute of Science and Technology, 08860 Castelldefels (Barcelona), Spain}
\affiliation{Universitat Polit\`{e}cnica de Catalunya, 08034 Barcelona, Spain}
\author{Carles Mili\'{a}n}
\email{carmien@upvnet.upv.es}
\affiliation{Institut Universitari de Matem\`{a}tica Pura i Aplicada, Universitat Polit\`{e}cnica de Val\`{e}ncia, 46022 Val\`{e}ncia, Spain}
%
%%%%%%%%%%%%
% ABSTRACT %
%%%%%%%%%%%%
%
\begin{abstract}
\setlength{\parindent}{-.14cm}\textbf{
Taming the instabilities inherent to many nonlinear optical phenomena is of paramount importance for modern photonics. In particular, the so-called snake instability is universally known to severely distort localized wave stripes, leading to the occurrence of transient, short-lived dynamical states that eventually decay. The phenomenon is ubiquitous in nonlinear science, from river meandering to superfluids, and to date it remains apparently uncontrollable. However, here we show that optical snake instabilities can be harnessed by a process that leads to the formation of stationary and robust two-dimensional zigzag states. We find that such new type of nonlinear waves exists in the hyperbolic regime of cylindrical micro-resonators and it naturally corresponds to two-dimensional frequency combs featuring spectral heterogeneity and intrinsic synchronization. We uncover the conditions of the existence of such spatiotemporal photonic snakes and confirm their remarkable robustness against perturbations. Our findings represent a new paradigm for frequency comb generation, thus opening the door to a whole range of applications in communications, metrology, and spectroscopy.
}
\end{abstract}
\maketitle
%%%%%%%%%%%%%
%%% FIG.1 %%%
%%%%%%%%%%%%%
\begin{figure*}[ht]
\centering
%\begin{center}
%%\includegraphics[width=.55\textwidth]{fig1a.png}
% EN UNA PART
%%\includegraphics[width=.99\textwidth]{FIG1_resonances2.pdf}
%%\includegraphics[width=.99\textwidth]{FIG1_resonances2_xicoteta.pdf}
% EN DOS PARTS
%%\includegraphics[width=.85\textwidth]{FIG1a.jpg}
%\includegraphics[width=.895\textwidth]{FIG1_resonances-2_IGS_cropped.jpg}
\includegraphics[width=.895\textwidth]{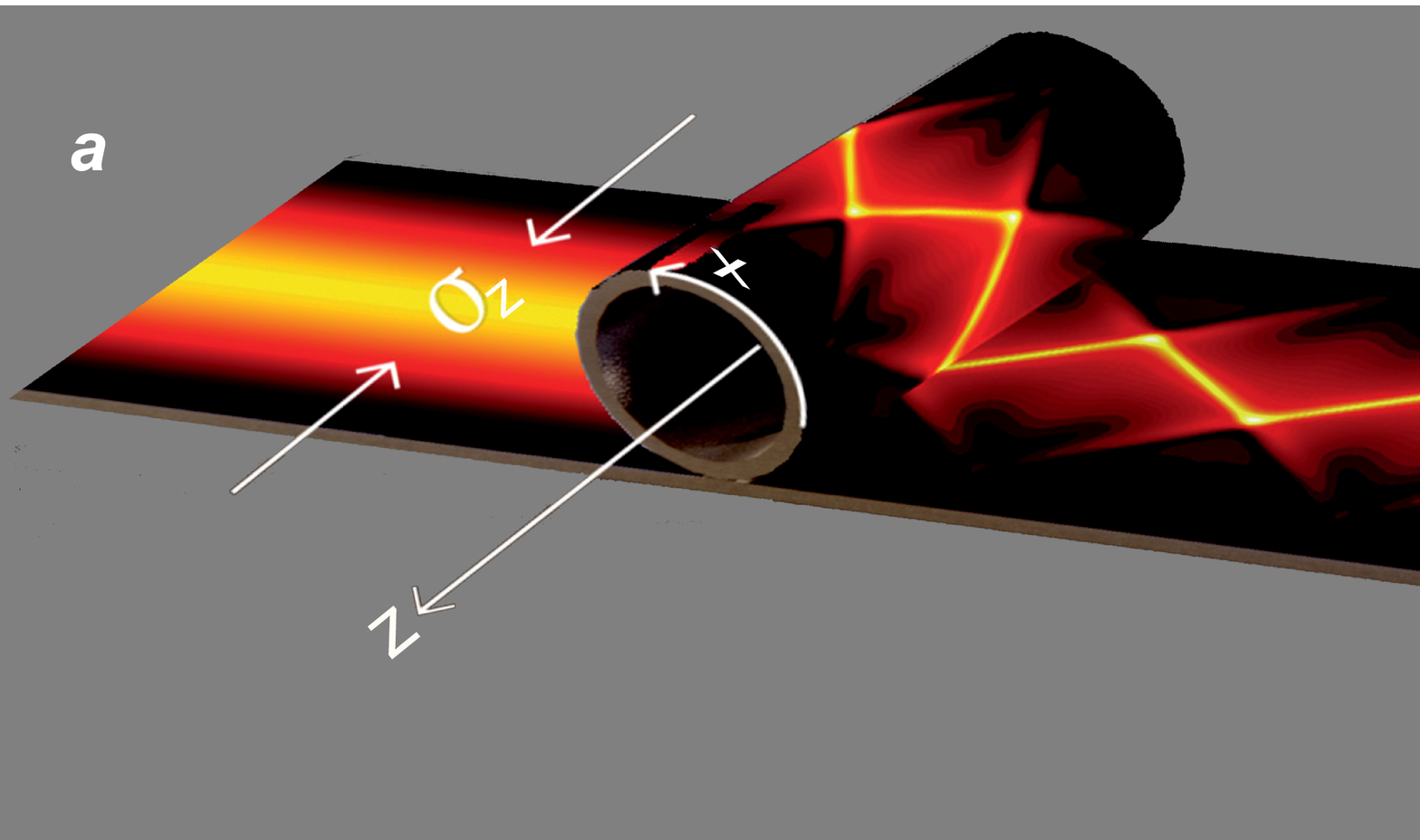}
\includegraphics[width=.99\textwidth]{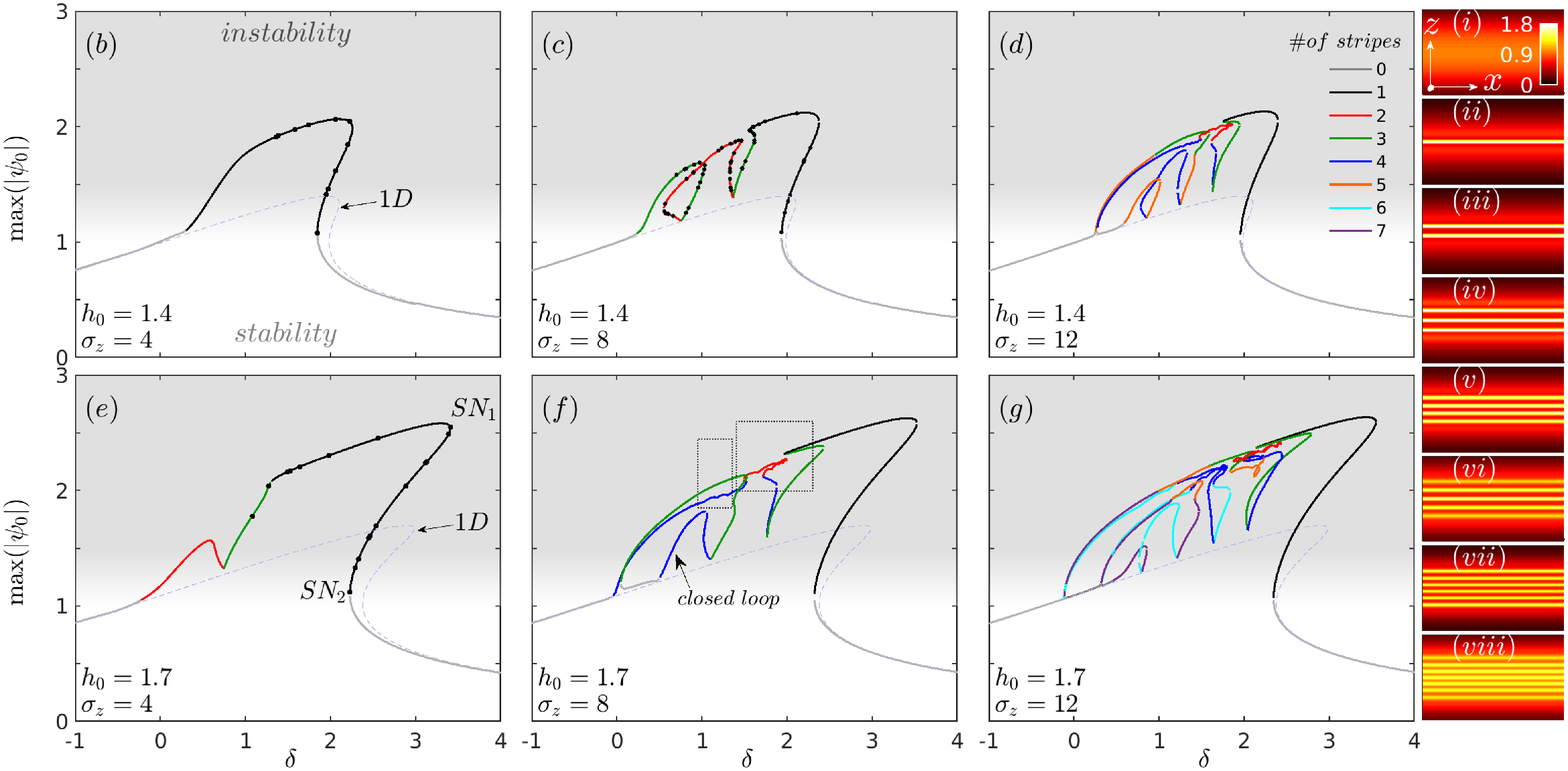}
\caption{\label{f1} \textbf{\textbar\ Cylindrical microresonator and multi-stable resonances. a}, sketch of the driven (from the left) micro-cylinder in the photonic snake regime. \textbf{b-g}, maximum of the cavity background (single color) field amplitude, $|\psi_0(x,z)|$, versus laser-cavity detuning, $\delta$, showing selected nonlinear resonances for driving strengths \textbf{b-d} $h_0=1.4$, \textbf{e-g} $h_0=1.7$, and various values of the transverse pump width: \textbf{b,e} $\sigma_z=4$; \textbf{c,f} $\sigma_z=8$; \textbf{d,g} $\sigma_z=12$. The cavity background may be smooth [inset $(i)$] or may feature single- and multi-stripe quasi one-dimensional solitons distributed along $z$ [insets $(ii)-(viii)$]. The number of stripes on each portion of the resonances is colour coded in the legend of \textbf{d}. The resonance corresponding to the \textit{one-dimensional} microring is also shown, for reference, by the dashed curves labeled as '$1D$' in \textbf{b}, \textbf{e}. Thick black dots on the resonances in \textbf{b},\textbf{c},\textbf{e} mark the onsets of instabilities and correspond to bifurcation points for spatiotemporal states, including snakes [see Fig.\ref{f2}]. Labels $SN_{1,2}$ in \textbf{e} mark the saddle node bifurcations (branch turning points) further discussed in Fig.\ref{f3}. Rectangles in \textbf{f} enclose the regions zoomed in Figs.\ref{f4}\textbf{b} and \ref{f4}\textbf{e}. Background states are stable for $|\psi_0|<1$ and may be unstable otherwise, within the grey shaded areas. Insets showing spatial profiles are plotted over the area $x\in[-L/2,L/2[$ (covering the whole cavity length, $L=16$) and $z\in[-16,16]$.}
%\end{center}
\end{figure*}
%
%%%%%%%%%%%%%
%%% FIG.2 %%%
%%%%%%%%%%%%%
\begin{figure*}[ht]
\centering
%\begin{center}
\includegraphics[width=.99\textwidth]{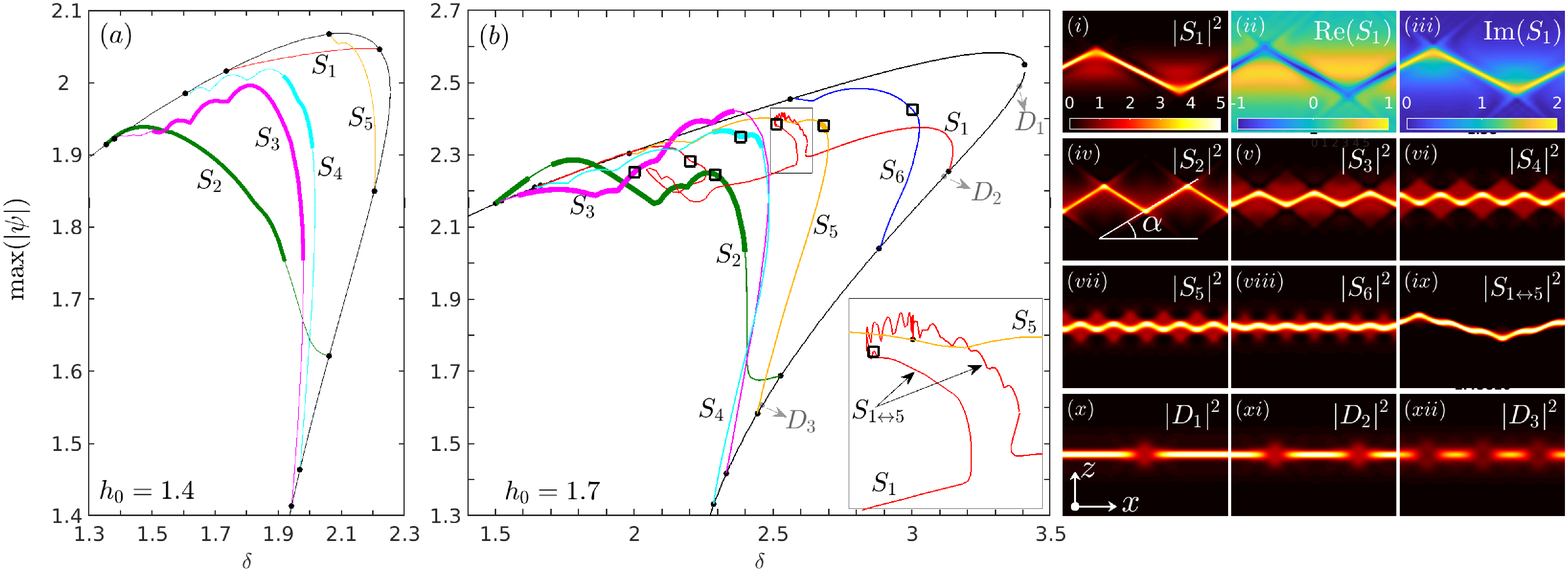}
\includegraphics[width=.99\textwidth]{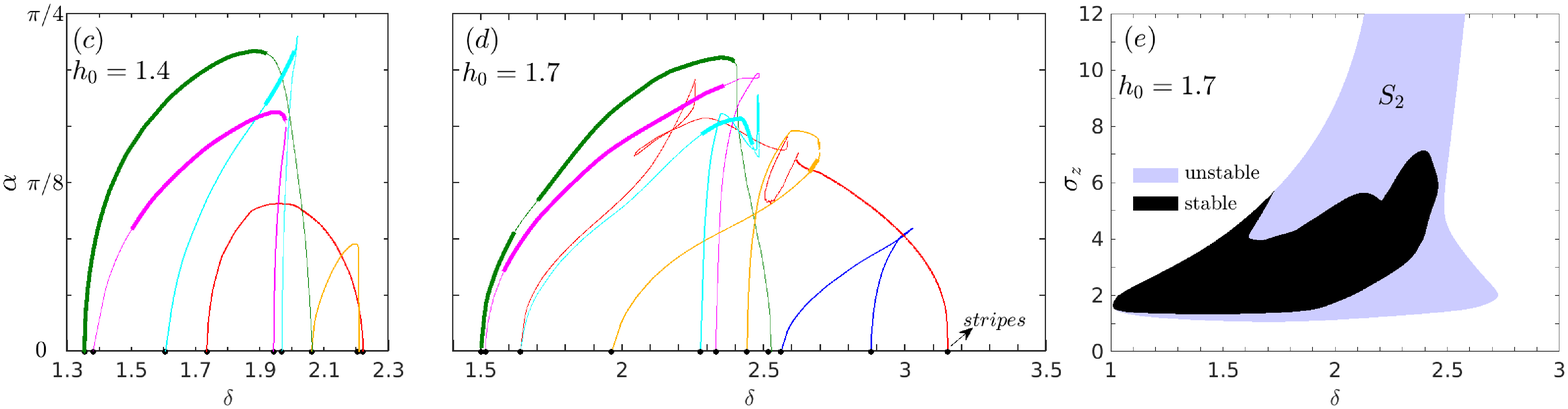}
\caption{\label{f2} \textbf{\textbar\ Photonic snake families bifurcating from the single stripe background. a,b}, amplitude vs detuning branches for all photonic snakes supported by the microcylinder considered in Fig.\ref{f1} with $\sigma_z=4$ for \textbf{a} $h_0=1.4$ and \textbf{b} $h_0=1.7$. Thick (thin) traces denote stable (unstable) states. The nonlinear resonances (black curves) are those shown in Figs.\ref{f1}b and \ref{f1}e, respectively. Photonic snakes exhibiting $N$ periods around the cavity circumference are denoted by $S_N$, and selected profiles, at the positions of the squares in \textbf{b}, are shown in insets $(i)-(viii)$. Panels $(ii),(iii)$ show real, imaginary parts of the snake $S_1$. Inset in \textbf{b} is a zoom over the region where the $S_1$ branch approaches the $S_5$ branch, and as a consequence, the corresponding snakes [inset ($ix$)] present a profile mixing both periodicities: 5 small zigzag periods within a 1 period envelope. Other branches exist (not shown), associated to $N$-dark soliton states, denoted by $D_N$, with typical profiles as shown in $(x)-(xii)$% (they are discussed in the supplementary information)
. \textbf{c,d} show the snake's zigzagging angle [cf. inset ($iv$)] as a function of detuning, $\alpha(\delta)$, corresponding to all branches in \textbf{a,b} (branches are colour-matched). \textbf{e} shows the existence (light area) and stability (black area) domains in the $\{\delta,\sigma_z\}$ plane for the snake family $S_2$ with $h_0=1.7$. All insets are plotted over the area $x\in[-L/2,L/2[$ and $z\in[-8,8]$.}
%\end{center}
\end{figure*}

%%%%%%%%%%%%%
%%% FIG.3%%%
%%%%%%%%%%%%%
\begin{figure*}
\begin{center}
\includegraphics[width=.99\textwidth]{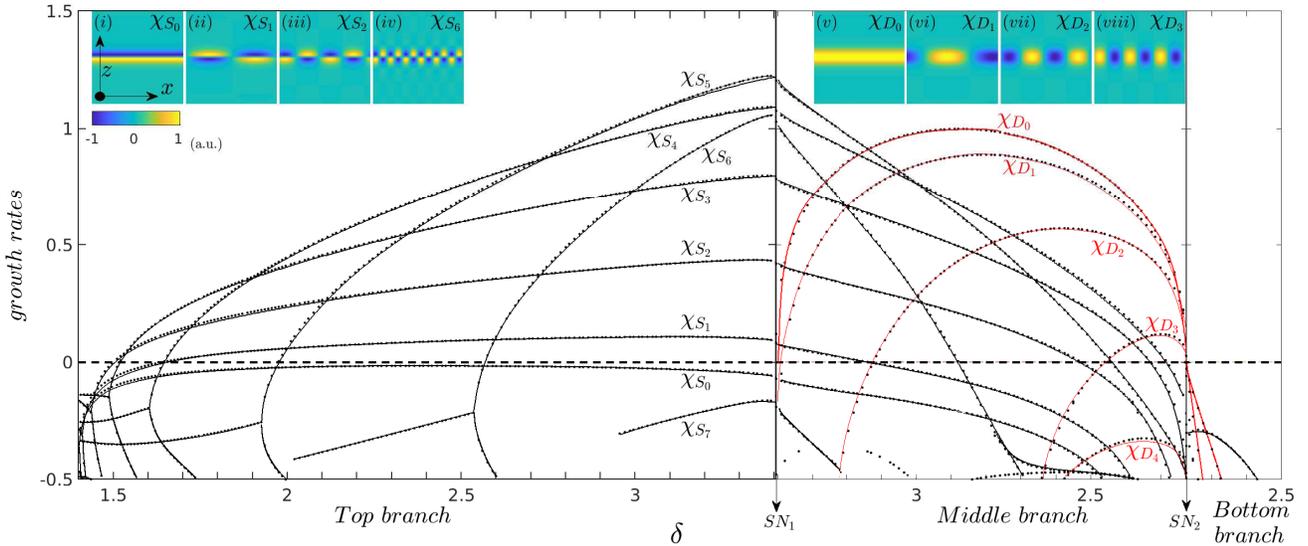}
\caption{\label{f3}\textbf{\textbar\ Onset of snakes.} Growth rates vs detuning of the internal modes associated to the background states defining the cavity resonance in Fig. \ref{f1}\textbf{e} [$\sigma_z=4$, $h_0=1.7$]. This figure is a parametric plot along the resonance, i.e., the abscissa corresponds to the values of detuning along the branches of the resonance, starting at $\delta=1.4$. The left panel ($\delta\in[1.4,SN_1\lesssim3.5$) corresponds to the top branch, the central panel ($\delta\in[SN_1,SN_2]$) corresponds to the middle branch, and the right panel ($\delta\in[SN_2\gtrsim2.2,2.5]$) corresponds to the lower and stable branch. Labels $\chi_{\psi}$ denote the internal modes the growth of which leads to the state $\psi$ [$\psi$ are either snakes $S_N$ (black) or dark solitons $D_N$ (red)]. Insets at the top show the real part of typical snake (left) and neck (right) modes. Snake modes with $N=0$ induce, when excited, a global drift along $z$ while neck modes with $N=0$ correspond to the \textit{universally} unstable mode that dotes the whole middle branch with the detrimental zero-wavelength instability. The zero growth rate points correspond to the bifurcation points shown in Figs. \ref{f1} and \ref{f2} from where the corresponding states (shown in Fig. \ref{f2}) emerge. Insets are plotted over the area $x\in[-L/2,L/2[$ and $z\in[-8,8].$}
\end{center}
\end{figure*}

%%%%%%%%%%%%%
%%% FIG.4 %%%
%%%%%%%%%%%%%
\begin{figure*}[ht]
\centering
%\begin{center}
\includegraphics[width=.99\textwidth]{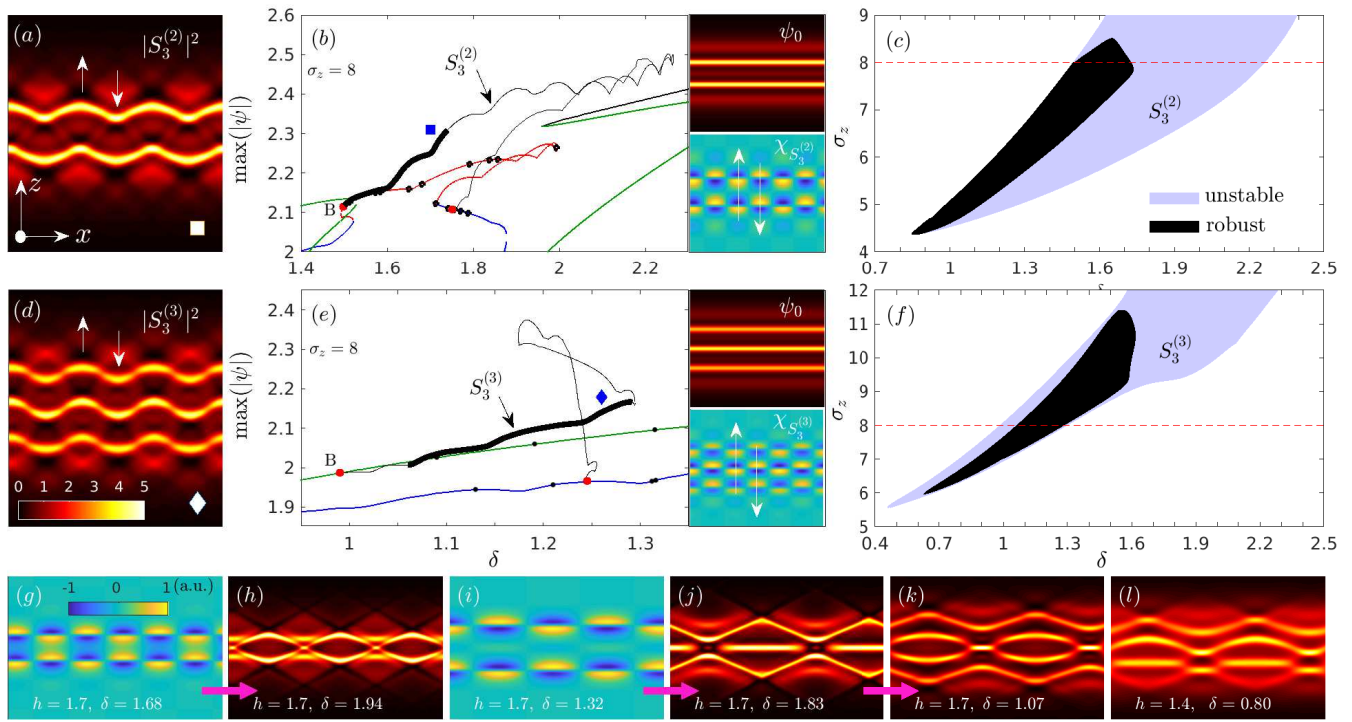}
\caption{\label{f4}\textbf{Coupled photonic snakes. a,} intensity of a 2-snake with 3 periods [denoted by $S_3^{(2)}$] for $h_0=1.7$ and $\sigma_z=8$. \textbf{b,} branches of its existence and stability shown by the solid and thick black curves, respectively. Branches with other colors correspond to the resonances in Fig. \ref{f1}f. A square marker is around the location where the profile in \textbf{a} is taken ($\delta=1.72$). The stable branch emerges from the bifurcation point labeled with B (left most thick red dot). Inset: (top) 2-stripe state, (bottom) perturbation impingin the flexural instability that transforms the 2-stripe state into the $S_3^{(2)}$ in \textbf{a}. White arrows indicate the direction of the local drift [arrows in inset and in \textbf{a} are in exact correspondence]. \textbf{c,} Existence and stability chart in the $\{\delta,\sigma_z\}$-plane for $h_0=1.7$, built by analysing the stability of the stable branch in \textbf{b} and varying pump-width. Horizontal dashed line marks $\sigma_z=8$, corresponding to \textbf{b}. \textbf{d-f} are analogous to \textbf{a-c} for a triple snake. Other snaking states, albeit unstable: \textbf{g,i} show the internal modes of the (2,3)-stripe states leading to states in \textbf{h,j}. Additional bifurcations reshape states of the type \textbf{j} into \textbf{k} and \textbf{l}. All $(x,z)$ panels are plotted over the area $x\in[-L/2,L/2[$ and $z\in[-10,10].$} 
%\end{center}
\end{figure*}
%
%%%%%%%%%%%%%
%%% FIG.5 %%%
%%%%%%%%%%%%%
\begin{figure*}[ht]
\centering
%\begin{center}
\includegraphics[width=.99\textwidth]{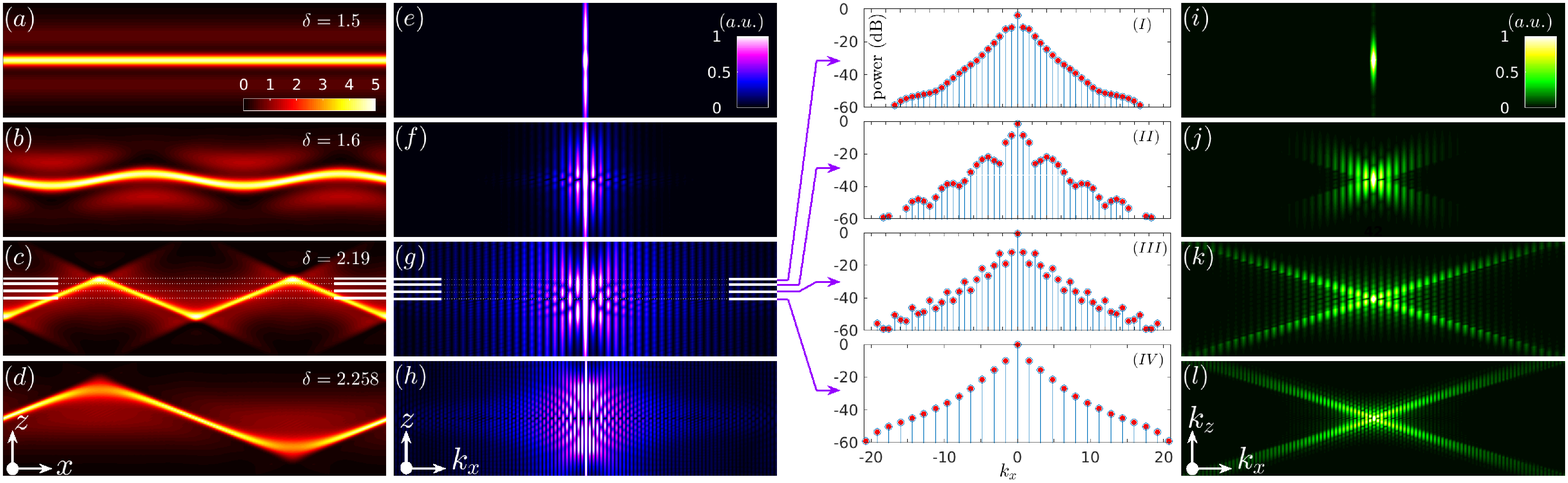}
\caption{\label{f5}\textbf{Two-dimensional heterogeneous combs. a-d,} intensity, $|\psi(x,z)|^2$, for several snake profiles with different tilt, $\alpha$, periodicity, and detuning, $\delta$ (see labels) at $\sigma_z=4$ and $h_0=1.7$. \textbf{e-h}, Fourier transform along the $x$ direction of the fields in \textbf{a-d}, illustrating the two-dimensional comb structure vs $z$ around the central wave-numbers, $k_x$. The frequency combs exhibit a high degree of heterogeneity along the cylinder's axis ($z$), as evident from insets {\scriptsize{(I)-(IV)}}, showing the broad one-dimensional combs at different locations: $z=1.72$, 1.25, 0.63, 0, respectively. \textbf{i-l}, Two-dimensional Fourier transforms of the fields in \textbf{a-d} illustrating snakes in momentum space, which reveal the angular spread of the two-dimensional comb frequencies (see text).  
Figure axes: \textbf{a-d}, $x\in[-L/2,L/2[$ and $z\in[-5,5]$; \textbf{e-h}, $k_x\in[-20,20]$ and $z\in[-5,5]$; \textbf{i-l}, $k_x\in[-20,20]$ and $k_z\in[-20,20]$. Spectra in \textbf{e-l} are normalised to the strongest comb line excluding the pump at $k_x=0$. 
}
%\end{center}
\end{figure*}
%
%%%%%%%%%%%%%%%%
% INTRODUCTION % MAKE IT 1.5 COLUMNS
%%%%%%%%%%%%%%%%

%\textbf{Introduction [remove this heading]}
%V2 - 1st paragraph
\lettrine{\textcolor{blue}S}{}ince their discovery almost half a century ago, the so-called flexural instabilities \cite{zakh_rub}, which lead to the unarrestable decay of quasi-one-dimensional self-sustained states such as solitons \cite{zakhPR,kivpel}, have been encountered in all areas of nonlinear sciences. Namely, they have been found to occur in the dynamics of river meandering \cite{meander}, classical fluids \cite{fluid}, fermion \cite{yefNat,cetoli} and polariton \cite{claude} superfluids, Bose-Einstein condensates \cite{bec}, chemistry \cite{chem}, and, importantly,  nonlinear photonics \cite{mam,tik,gorza04,gorza11,gorza06}. The initial stages of the instability $-$ referred to as the \textit{snake} instability \cite{zakhPR}$-$ evinces the reshaping of straight states into zigzags caused by spontaneous and local transverse drifts with alternating directions, which occur prior to the irreversible decay of the corresponding states. In contrast to the use of the instability in several systems, e.g., to trigger spontaneous vortex formation \cite{bec,mam,tik}, its controlled arrest to form robust  multi-dimensional snaking states remains hitherto unexpected.
%V1 - 1st paragraph
%\lettrine{\textcolor{blue}S}{}ince their first discovery almost half a century ago, flexural instabilities \cite{zakh_rub} leading to unarrestable decay of quasi-one-dimensional self-sustained states, such as solitons \cite{zakhPR,kivpel}, have been encountered in diverse physical systems. The onset of this instability is characterised by the reshaping of straight states into zigzagging forms, as a consequence of spontaneous transverse drift with alternating directions. This instability $-$the \textit{snake} instability \cite{zakhPR}$-$ is found over the whole realm of nonlinear sciences and it has been unambiguously identified in the dynamics of river meandering \cite{meander}, classical fluids \cite{fluid}, fermion \cite{yefNat,cetoli} and polariton \cite{claude} superfluids, Bose-Einstein condensates \cite{bec}, chemistry \cite{chem}, and nonlinear optics \cite{mam,tik,gorza04,gorza11,gorza06}. So far, this instability has been of great fundamental interest and also very useful for triggering complex dynamical transformations such as spontaneous formation of vortices \cite{bec,mam,tik}. However, given its omnipresent destructive character, its control to form stabilised multi-dimensional snake states remains a highly unexpected fundamental phenomenon.

In parallel, in recent years, microring cavities have emerged as an outstanding platform for exploring the fundamental properties and the applications of stabilised one-dimensional dissipative nonlinear waves, in both the normal and the anomalous dispersion regimes, under a plethora of potentially unfavourable situations, such as strong dispersive \cite{braschSci,vahNatCom}, inelastic \cite{karpov,VahalaStokes,gaeta}, and thermal \cite{stone,obrther} effects. Each and every new solitonic states found in such micron-sized systems correspond to highly stable frequency combs that contribute to the developing field of comb-related applications \cite{pasqPR,cundiffRMP}. As a consequence, microring-like cavities are highly appealing systems to push the frontiers of knowledge of nonlinear phenomena as well as boosting applications in photonics. Nowadays, most solitonic states (stable combs) that have been experimentally observed in microrings occur in essentially single-comb forming devices, except for the few exceptions that may be found in counter-propagating soliton experiments \cite{vahCount,kippCount}, in systems with two microrings \cite{dimerKipp,gaetfew}, or in cavities with a few transverse modes \cite{kippsdm}, where each of the individual combs are formally one-dimensional. The potential of two-dimensional frequency combs, which to date is essentially unexplored experimentally, remains to be uncovered.

Here we discover the existence of robust spatiotemporal photonic snakes in cylindrical micro-resonators with normal group velocity dispersion (GVD) and in the presence of diffraction along the cylinder's axis. Such previously unknown states are a continuous two-dimensional ensemble of heterogeneous combs which are inherently synchronised by the nonlinearity. Thus, they represent a whole new paradigm for frequency comb formation.

%[WHERE WILL YOU DISCUSS THE ADVANTAGES OF THE NORMAL GVD??? THIS IS IMPORTANT AND MAKES ME STEP AWAY FROM ANOMALOUS GVD WORKS, INCLUSING MY OWN PRL's!!. EITHER IN INTRO OR IN DISCUSSION, OR A BIT GLIMPSE IN INTRO AND FURTHER ELABORTAED IN DISCUSSION.]

% CHECK IF YOU NEED REFS FROM COMMENTED PARAGRAPH BELOW:
%The development of microcomb sources \cite{vahala03} opened new prospects that are deeply impacting several fundamental and applied research areas 
%\cite{vahVern,vahCount,suh16,papp14,spen18,marinpal17,pfeifle14,suh18,Bao19,ben10,ozb19} (see Refs.\cite{pasqPR,cundiffRMP} for reviews). One of the most transcendental achievements was the demonstration of a frequency comb carried by a rigorous soliton in a Kerr micro-cavity \cite{herr14} and its stabilization \cite{yi16,brasch16,guo17}. 

% POSSIBLE REFS ?

% \cite{cetoli} good one about fermionic superluids, theory, try tio cite it!

%new form of stable comb, namely two-dimensional frequency combs which may be regarded as a wide heterogeneous collection of perfectly synchronised and individually accessible combs. Hence, our results represent a novel paradigm of frequency comb formation which are expected to offer unique advantages for some applications as well as inspire new routes in the field of frequency combs.

%%%%%%%%%%%
% RESULTS %
%%%%%%%%%%%
\ 

\noindent\textbf{\large{Results}}

% One opening paragraph before going into subsections:

% MEssage: WE FOUND ROBUST 2D-PHOTONIC SNAKE STATES, and THEY carry heterogeneous 2D COMBS}

% EMPHASISE: "unexpected, fundamentally new, solidly predicted, and new paradigm with realistic implementation".

%to form two-dimensional photonic snake states. We achieve this unprecedented control 

% The snakes orbit around the hollow micro-cylinder, confined within its thin wall (hosting only one radial mode)

\begin{figure}%[ht]
\centering
%\begin{center}
\includegraphics[width=.49\textwidth]{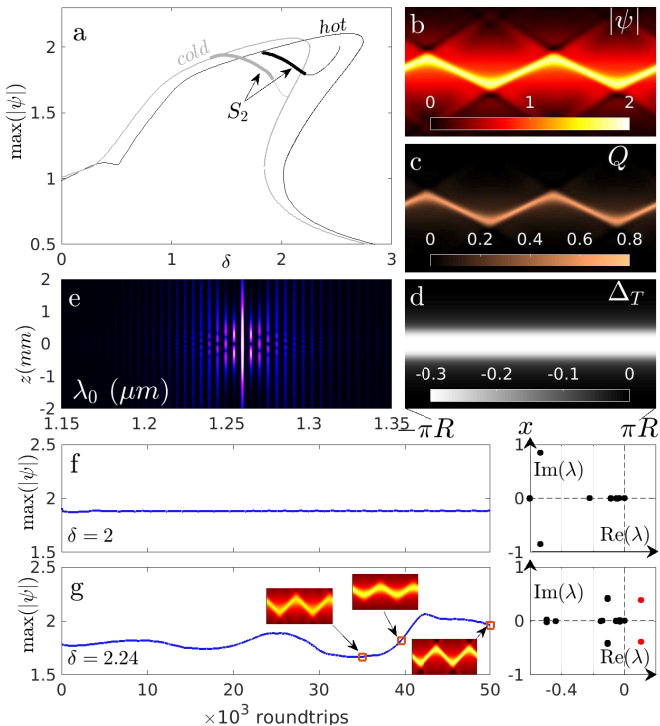}
\caption{\label{f6}\textbf{Effect of perturbations. a,} cavity resonance and the branch for the two-period snakes, $S_2$, affected by perturbation originating from higher order dispersion, Raman scattering and thermal nonlinearity (black) [resonance is solely affected by thermal effects, and hence the label \textit{hot}]. The results without perturbations are also shown for reference (gray, cf. Fig.\ref{f2}\textbf{a}). \textbf{b-d} show, respectively, the light field, the Raman vibration, and the thermal distribution for the stable snake at $\delta=2$. \textbf{e}, the corresponding two-dimensional comb. \textbf{f}, the maximum amplitude of the light field during the propagation of the stable and robust snake ($\delta=2$), for a total time of $155$ ns [50,000 roundtrips], under initial noise and perturbations at each step. Right-inset: complex eigenvalues, $\lambda$, obtained from the stability analysis performed on the stationary snake used as input of the propagation, taking into account all above mentioned perturbations. Instabilities correspond to positive real parts of $\lambda$. \textbf{g}, analogous results as for \textbf{f}, but for an oscillatory unstable snake ($\delta=2.24$). Small insets illustrate the dynamical reshaping of the snake with time. Panels \textbf{b}-\textbf{d} and insets in \textbf{g} span over the whole cavity circumference and over 4 mm along $z$.}
%\end{center}
\end{figure}

\noindent
\noindent\textbf{Model.} The spatiotemporal snakes predicted here form in hollow cylindrical Kerr micro-resonators [cf. Fig. \ref{f1}a]. The dynamics of the intra-cavity field can be accurately described by the two-dimensional generalisation of the Lugiato-Lefever equation \cite{lugiato87},

\begin{equation}
\partial_t\psi= \frac{i}{2}(-\partial_x^2+\partial_z^2)\psi
-(1+i\delta)\psi+i|\psi|^2\psi+ih_0e^{-z^2/\sigma_z^2},
\label{eq1}
\end{equation}
where the terms on the right-hand side of the equation account for normal GVD along the propagation (and periodic) coordinate, diffraction along the cylinder's axis, losses, laser-cavity detuning, focusing nonlinearity, and the axially localised pump, respectively (see methods). For the sake of generality, we discuss photonic snakes first in the physical frame of Eq.\ref{eq1} [cf. Figs. \ref{f1}-\ref{f5}], which represents a universal hyperbolic model governing a plethora of nonlinear physical phenomena in optics \cite{op1,op2} and other areas of physics \cite{kevrev}. %\cite{ablowitzsegur,w1,w3,per,myra,lit}.
%
%phenomena such as capilary waves \cite{ablowitzsegur}, water waves \cite{w1,w3}, plasmas [17,18,13*,12*], and optics \cite{op1,op2} (see \cite{kevrev} for a review). %[SEE REFS 1,10-20 in kevrekides' review and complement/substitute the above list].
%
In order to anchor our predictions into realistic optical frequency comb forming experimental realizations, we report on the snakes' robustness under the specific collection of perturbations expected in cylindrical micro-cavities [cf. Fig. \ref{f6}], including Raman and thermal nonlinearities, among others (see methods).

\ 

\noindent\textbf{Nonlinear resonances.} When driven into the strongly nonlinear regime, the here considered micro-cylinders
%present tilted cavity resonances  which may lead to bistability as the system has, at most, three possible single-color (background) states for a fixed detuning (albeit one is fundamentally unstable). However, in 
%
%the interplay between normal dispersion and diffraction, being the latter formally equivalent to anoamlous GVD along $z$, results in exotic
%
develop nonlinear resonances, associated to the single colour locked states, $\psi_0$ ($\partial_t\psi_0=\partial_x\psi_0=0$), exhibiting an unusually rich multi-stability landscape, as shown in Figs.\ref{f1}b-g. For peak intensities $|\psi_0|\gtrsim1$, corresponding to the existence domain of bright solitons with anomalous GVD \cite{BarashPRE96} (formally analogous to diffraction along $z$), the background states become single and multi-stripe solitons [cf. insets $(ii)$-$(viii)$] distributed on top of the smooth background [cf. inset $(i)$], recalling the spatial solitons reported in pioneering works \cite{mclaPRL83}. The presence of stripes impinges an intricate morphology of the resonances strongly deviating from that of the tri-valued (bistable) one-dimensional counterparts (marked for reference by the light-grey curves in Figs.\ref{f1}b-g). Resonances associated to wider or more intense pumps (i.e., greater $\sigma_z$ and $h_0$, respectively) contain states hosting more stripes and reshape to exhibit splitting (Figs. \ref{f1}c,e), nesting (Figs. \ref{f1}d,f,g), and closed loops (Fig. \ref{f1}f). As discussed below, the properties of resonances are intimately linked to the existence of snake states.

\noindent\textbf{Photonic snakes.} The soliton stripes described above are prone to snake instabilities. When these develop, stripes distort and acquire in our dissipative system a periodic zigzag snaking profile, becoming inhomogeneous along $x$ and, hence, polychromatic. The central result of this work is that the transverse drifts induced by the snake instability are fully arrested after they exerted certain distortion to stripes so that perfectly stationary snakes form [see Figs. \ref{f2}$(i)$-$(ix)$ for typical profiles]. Figures \ref{f2}a, b present the branches [$\max(|\psi(x,z)|)$ vs detuning] corresponding to the complete set of snakes existing with narrow pump, $\sigma_z=4$, for different driving strengths, $h_0$ (see caption). Snakes featuring from 1 to 5 periods in the microcavity circumference exist and those with 2 to 4 periods are stable for some detuning intervals. We denote different branches of snakes featuring $N$ periods on the microresonator circumference as $S_N$. %From the representative snake profiles shown in insets $(i)-(ix)$, we observe that the zigzags do not take arbitrarily large angles with respect to the propagation direction, $\alpha$. 
Figures \ref{f2}c,d show that snakes feature pronounced zigzagging angles, $\alpha$ [cf. inset $(iv)$], over most of their existence regions, so that they have a strong trend to form narrow pulses along $x$. %We estimated that $\alpha$ should not exceed the value $\pi/4$ (see supplemental material), consistent with all our observations.
We emphasize that the robust snakes exist over a finite and generous region of the parameter space, as shown in Figs. \ref{f2}a-d. To further illustrate this, Fig. \ref{f2}e shows the existence domain on the $\{\delta,\sigma_z\}$ plane for the $S_2$ family, which is found to be stable within the black area, unstable within the light shaded area, and non-existent otherwise (white area). In this work, optimal conditions for the snakes stability are encountered only for strong pump localisation along $z$. Outside the stability domain, snakes are exposed to oscillatory and exponential instabilities leading, respectively, to breathing and decay. The precise combination of the driving amplitude and aspect ratio of the pump field, $\sigma_z/L$, (lying in the range $[\frac{1}{8},\frac{1}{3}]$ in Fig.\ref{f2}e) is essential for the formation of robust snakes. 
%Consistently, important experimental studies \cite{yefNat} report on different snake instability time-scales upon the adjustment of the system's aspect ratio.

%We expect that our findings will inspire the frequency comb community as well as many others where the snake instability has been observed. but the rare states we are reporting here where never successfully formed (benefit for those communities?? any at all? control fluid mixtures, BECs, probably superfluids [Nature paper] a clear example). THIS IS in the end a discussion on the origin of stationary/stable snakes, essential in this work! Perhaps the dissipative nature f the system is also relevant.

%Try to give some numbers in other systems, relate at least the width of the snake with that of the background, perhaps from various papers you can get this. Perhaps this should go into discussions! but it is important! and then show a lot of biblio, that you know it and that you bloody know what you are talking about. This discussion is actually the main value of this discussion.

%In addition to snakes, we also found dark solitons [cf. Fig.\ref{f2}$(x)-(xii)$] emerging from stripes via neck instabilities at the bifurcations marked in Fig.\ref{f2}b with grey dots, labeled $D_1$, $D_2$, $D_3$. All dark solitons found by us were unstable (they are further discussed in the supplementary information).

% points from where they emerge are discussed below in FIg.3!

% PREVIOUS SECTIONING -ONSET OF SNAKES-

%\ 

%\noindent
%\textbf{Onset of photonic snakes.}
%

Photonic snake families emerge supercritically from the top branch of the cavity resonance, each with a different detuning threshold (marked with black dots in Fig.\ref{f2}), corresponding to the detuning values at which stripes become unstable due to the growth of snake type perturbations or \textit{internal modes} \cite{internal2} %\cite{internal1,internal2}
(see methods for the stability analysis description). Figure \ref{f3} shows all relevant growth rates for the three branches (top, middle, and bottom) of the right-most cavity resonance shown in Fig. \ref{f1}e. Snake-type perturbations ($\chi_{S}$, see top insets in Fig. \ref{f3}) with different periods acquire positive growth rates on the top and middle branches of the resonance. The nullity of growth rates defines exactly the bifurcation points (black dots in Fig. \ref{f2}b) or loci from which snakes emerge from stripes. We note that the symmetry properties of the stripes internal modes allows one to readily predict the symmetries to be inherited by the emerging snake family.

%WE MAY GET RID OF THIS POINT BELOW 
%%%%%%%%%%%%%
% ADD or REMOVE??
%We emphasize that snake instability generally exists over a region of the parameter space substantially larger than that were snakes are found as stationary states. As an example, Fig.\ref{f3} shows that a mode $\chi_{S_2}$ has positive growth in the range of $\delta\approx [1.5,3.4]$ while the snakes $S_2$ exist only within $\delta\approx [1.5,2.5]$, thus the presence of the instability does not guarantee the existence of snakes, stable or unstable.

Although our attention is focused on the snake instability, we point out that the middle branch of the resonance in Fig. \ref{f3} also presents \textit{neck} instabilities \cite{zakh_rub}, which often compete with the snake type \cite{skrPRL98}%\cite{skrPRL98,skr99,skr99b}
and manifests upon the growth of the corresponding internal modes, denoted by $\chi_{D_1}$-$\chi_{D_3}$ and shown in the right inset of Fig. \ref{f3}. In this work, all of the spatiotemporal states found with traces of neck perturbations, namely, the dark solitons in Figs. \ref{f2}($x$-$xii$) and the hybrid snake-dark solitons in Fig. \ref{f4}j,l, were highly unstable and thus their properties are not discussed in details.% (the existence and impact of neck instabilities in this system are discussed in the supplementary information).

%\noindent\textbf{Coupled snakes.}
The photonic snake families become much richer when snake instabilities develop on the multi-stripe states present on cavity resonances with large pump beam widths, $\sigma_z\gtrsim8$ (cf. Fig. \ref{f1}), what leads to the formation of coupled multi-snake states. Figure \ref{f4}a shows an example of robust two-snake state with three periods along the microcylinder's circunference, denoted by $S_3^{(2)}$. This double snake forms when the snake-type mode $\chi_{S_3^{(2)}}$ (Fig. \ref{f4}b bottom-right inset) grows on top of the two-stripe state (Fig. \ref{f4}b top-right inset). The full amplitude vs detuning branch of the $S_3^{(2)}$ snake is shown in Fig. \ref{f4}b by the black curve. Other branches correspond to the cavity resonance (cf. Fig. \ref{f1}f). Existence and stability of the $S_3^{(2)}$ family with $h_0=1.7$ is shown vs pump width and detuning over the relevant parameter space in Fig. \ref{f4}c. Figures \ref{f4}d-f show analogous results for a triple-snake. The above two snake families correspond to particular cases where their branches connect background states with different number of stripes: the $S_3^{(2)}$ [$S_3^{(3)}$] family bifurcates at lower $\delta$ from the $\mathrm{B}$-point located on a 2 (3) stripe state and merges with 4-stripe state at a higher $\delta$ (red dots).

Even though we only found stable \textit{in-phase} multi-snakes, the system supports a vast collection of states with different morphology. Some of these are illustrated in Fig. \ref{f4}h and Figs. \ref{f4}j,k,l which appear after the \textit{anti-phase} snake modes in Figs. \ref{f4}g and \ref{f4}i grow, respectively, on two- and three-stripe states. This results in the formation of anti-phase snakes that appear alone (Fig. \ref{f4}h), mixed with dark solitons (Fig. \ref{f4}j), together with ellipsoids  (Fig. \ref{f4}k), or displaying asymmetries along $z$ (Fig. \ref{f4}l).

\ 

\noindent\textbf{Heterogeneous 2D combs.}
The most remarkable feature of photonic snake states is that their spectrum is heterogeneous along $z$, while being inherently synchronized by the nonlinearity. Figure \ref{f5} shows a soliton stripe (Fig. \ref{f5}a) and snakes of different tilts and periods (Figs. \ref{f5}b-d), together with their frequency comb distribution along $z$ (Figs. \ref{f5}e-h). %, obtained by the Fourier transform of the snakes along the cylinder's circumference, $x$: $\hat{\mathcal{F}}_x\{\psi(x,z)\}(k_x,z)$.
Heterogeneity is evident in Figs. \ref{f5}g,h, showing Fourier spectra along the $z$-axis, and it is further illustrated in insets $I-IV$ by showing different combs extracted at different axial positions (marked by horizontal white lines in Figs. \ref{f5}c,g). Synchronisation and heterogeneity of these spectra is readily important for metrology and spectroscopy \cite{picqueNP}%, and may one day enable the concept of multi-comb spectroscopy
. The spreading of the combs along the cylinder's axis, $z$, naturally introduces the notion of heterogeneous two-dimensional comb, which constitutes a generalisation of the widely reported one-dimensional comb and a central result of this work.

Figures \ref{f5}i-l show the snakes in Figs. \ref{f5}a-d, respectively, in the two-dimensional momentum space, illustrating the angular spread of the different spectral components, potentially important for an efficient collection of the combs by external tapers or waveguides. Because dispersion in cylinders is typically much smaller in $x$ than in $z$ \cite{gorodjosab,milianPRL18}, the physical values corresponding to $k_z$ are in practice much smaller than those corresponding to $k_x$, so that the propagation angles $\theta=\arctan({k_z/k_x})$ remain in the order of a few degrees, at most (see methods). % between [-2, 2] approx.
On the same reason, the height of panels $e-h$ is of the order of a few mm's, so that the comb heterogeneity occurs along $z$ over larger scale than typical taper fiber widths, which enables the efficient light collection by, e.g., arrays of waveguides.

%
%%%%%%%%%%%%%
%%% FIG.6 %%%
%%%%%%%%%%%%%

\ 

\noindent
\textbf{Robustness.}
The photonic snakes and the two-dimensional combs exist in a physical setting that is readily realizable. In particular, consider a hollow micro-cylinder made of silica glass, with a radius $R=100\ \mu$m and wall thickness $w=0.75\ \mu$m, pumped at $\lambda_p\approx1.26\ \mu$m, with a $\mathcal{Q}$-factor of $\sim5\times10^6$, and take into account the expected specific effects introduced by chromatic dispersion, Raman and thermal nonlinearities, and the localised coupling, in $x$ and $z$, between the pump beam and the micro-cavity (see the full model in methods). Although each of the above \textit{perturbations} may be suppressed in various ways, we now show that robust snakes occur in passive driven cylindrical micro-resonators. Figure \ref{f6} presents an overview of our predictions in such scenarios, focusing the attention on the snake family $S_2$ with $h_0=1.4$ and $\sigma_z=4$. Figure \ref{f6}a shows the thermal shift of the nonlinear resonance [black] with respect to the unperturbed cold resonance [grey], as well as the $S_2$ snake branch, which stable region is highlighted by the thick trace. An example of robust snake is shown in Fig.\ref{f6}b, along with its associated Raman vibration (Fig.\ref{f6}c) and thermal (Fig.\ref{f6}d) fields. The corresponding two-dimensional comb is shown around the central lines in Fig.\ref{f6}e. We emphasize that in the context of one-dimensional combs, solitons and related nonlinear waves have been reported under the thermal nonlinearity \cite{iltherm}, both experimentally \cite{stone} and theoretically \cite{menyukther}. In one dimension, thermal effects mainly induce a global shift in the laser-cavity detuning which, even if they introduce their own instabilities \cite{iltherm,menyukther}, cannot compromise soliton existence itself. However, in the two-dimensional case here considered, thermal effects are inhomogenous along $z$ (cf. Fig.\ref{f6}d), and hence induce \textit{thermal stress} along $z$. As a consequence, the mere existence of the two-dimensional snakes cannot be anticipated \textit{a priori}, and neither can their stability. The latter is formally predicted via linear stability analysis and explicitly checked by long propagation runs (cf. Fig.\ref{f6}f) spanning over $50,000$ cavity roundtrips (equivalent to 155 ns in our geometry). Propagation of robust snakes under the whole plethora of realistic effects feature constant peak amplitudes (as a consequence of the steady two-dimensional profiles) despite the input random noise and step-to-step perturbations. On the contrary, unstable snakes behave very differently. In particular, Fig.\ref{f6}g shows the time evolution of an oscillatory unstable snake, featuring peak amplitude and profile variations (see insets). The presence and absence of instabilities, and the nature of them in the former case, are accurately predicted by the stability analysis (see methods), which results are displayed in insets of Figs.\ref{f6}f,g. %As a result of this work,  %which results are displayed in  the positive growth rates of the snake's internal modes, which are associated to the eigenvalues of the stability analysis, $\lambda$, marked in red in the right-most inset of Fig.\ref{f6}g (see methods).

\ 

\noindent\textbf{\large{Discussion.}}

\noindent
% EASY EXCITATION IN ANY SPECTRAL WINDOW
Stable photonic snakes present a trend to bifurcate super-critically from the stable background upon the increase of detuning [cf. Figs. \ref{f2},\ref{f4}], even in the presence of \textit{higher order effects} [cf. Fig. \ref{f6}], what strongly suggests that they will be easily and deterministicaly excitable in realistic experiments by the standard dynamical red-shift of the pump’s frequency. %even in the presence of thermal effects which play a major role in the dynamics of comb excitation [active capture, thermal Herr].
This simple excitation mechanism of frequency combs in the normal GVD regime of passively driven cavities, naturally provided here by the intrinsically two-dimensional snakes, has remained hidden till date by the low dimensionality of microrings, where relatively complex excitation methods were typically required. Indeed, in one-dimensional microrings, solitonic combs in the normal GVD regime are often excited by engineering a small anomalous GVD region around the pump’s frequency, so modulational instability triggers comb formation. This is achieved via mode coupling effects arising due to avoided crossings in multi-modal microrings \cite{saavOE12,XueNP,jang16,nazem21} or with the aid of an auxiliary ring \cite{xuebis,kim}. A recent work demonstrates a much simpler and straight-forward mechanism to excite (one-dimensional) dark solitons via self-injection-locking \cite{VahalaSILDarlSol} and displaying the turnkey \cite{turnkey} operation. Photonic snake states may thus belong to the collection of simple and deterministically excitable combs in cavities with dominant normal GVD, which is of central importance for extending the formation of micro-cavity combs to the long and short wavelength regions, far away from the telecom band around $1.5\ \mu$m.%, which is of current great interest [CITE].

% HETEROGENEITY AND SYNCHRONISATION OF TWO-DIMENSIONAL COMBS
In addition, the two-dimensional combs reported here automatically enable the possibility to host synchronised heterogeneous combs in a single device, features that are of great importance in metrology and spectroscopy, and may be difficult to combine in this specific manner with microrings. Indeed, comb heterogeneity may be achieved in the one-dimensional context, e.g., via the excitation of unbounded solitons copropagating along the same spatial channel \cite{heteroNC}. However, their different group velocities unavoidably yield de-synchronisation. On the other hand, synchronisation was previously reported for identical combs in distant microrings \cite{gaetSync}. Remarkably, both features were found simultaneously in the bi-modal stokes solitons \cite{VahalaStokes}, where heterogeneity appeared as a result of two copropagating and spectrally non-overlapping combs. In the present work, differently with the above, and akin to the two-dimensional geometry, the heterogeneity appears as a continuous comb reshaping along the cylinder’s axis and with a fixed carrier frequency. We note that the above remarkable findings \cite{heteroNC,gaetSync,VahalaStokes} were reported in the anomalous GVD region, where the physics is substantially different than that in the normal GVD regime, subject of this work.
% CAREFUL! Your cluster-PRL is also hetero+sync, so not novel. Even with microrings, Vahala stokes is also a two-mode comb, sync+hetero, so... mention that these are important features but do not make this your strong point. Your point is that you CONTROL an OLD INSTABILITY to form NEW STATES, in turn THEY form 2D combs, exhibiting a continuous transformation along z (heterogeneity is continuous) and that they occur in the normal GVD region, and also that excitation seems to be the simplest one found so far in normal GVD (such simple exciattion resembles that of turing rolls in 1D and anomalous because of the supercritical bif.). 

In summary, we have uncovered a fundamentally new mechanism to arrest and control the ubiquitous and otherwise strong snake instability in cylindrical Kerr micro-resonators and, as a result, the possibility to form complex robust spatiotemporal snake states. Specifically, we have found that photonic snakes naturally form perfectly synchronised heterogeneous comb ensembles in the normal GVD (hyperbolic) regime of cylindrical micro-cavities. The phenomenon represents a novel paradigm for the generation of optical frequency combs.

% DISCLAIMER: IF SOMEONE ASKS ABOUT OUR PREVIOUS PRL'S
%These states represent robust heterogeneous multi-frequency comb states. Results presented here are qualitatively very different from previous multi frequency comb states, which were all copies of one another \cite{IvarsPRL2021}. In the current work, the multi combs are made of a collection of combs with very different spectral features, such as the comb line spacing. Remarkably, all combs are perfectly synchronized despite their heterogeneity. The possibility to host many and very different combs in one single device has never put forward before.

% DISCLAIMER: SNAKE PROPOSED IN MATHEMATICAL MODELS, BUT NO STABILITY
%Last, we note that the existence of stationary snake states has only been suggested formally, to the very best of our knowledge, in mathematical models such as the double-sine Gordon \cite{linDSG} equations, often related to distant topic of crystal dislocations. Crucially, no allusion to their stability was made in any way. Hence, the formation of robust snake states remained, till date, a fully open problem. 

% A THOUGHT TO ENHANCE POTENTIAL IMPORTANCE
%You may link heterogeneity of combs with, up to certain extent, arbitrary waveform generation we thus achieve certain degree of arbitrariness on the waveform along the cylinder's axis, what can be of big practical interest \cite{cunNP}. [NOT SURE THOUGH]

\ 

\noindent\textbf{\large{Online content}}

\noindent
Methods, additional references, statements of data availability, acknowledgements, details of author contributions and competing interests are available in the online version of the paper.

\clearpage

%%%%%%%%%%%%%%%%%%%%
%% ONLINE METHODS %%
%%%%%%%%%%%%%%%%%%%%

\noindent\textbf{\large{Online Methods}}

\ 

\noindent\textbf{{Time evolution equations.}} 
The nonlinear dynamics of the electric field's envelope orbiting around a cylindrical microresonator is described by the following system of coupled equations,
%
%\begin{widetext}
\begin{eqnarray}
&& \nonumber i\partial_t\psi= -ib_{1,1}\partial_X\psi+\frac{1}{2}(\partial_X^2-\partial_z^2)\psi+\hat{D}_{hod}\psi
-[i-\delta]\psi-\\ &&-[(1-f_R)|\psi|^2-\Delta_T+Q]\psi-h_0\exp\left(-\frac{z^2}{\sigma_z^2}\right)\xi(X)\label{eq2},
\\ &&
\partial_t\Delta_T=-A\int_0^L|\psi|^2\frac{dX}{L}-B\Delta_T,\label{eq3}\ \ \\ &&
\partial_t^2Q=-\frac{2\gamma_R\tau}{\gamma}\partial_tQ-\frac{\tau^2\Omega_R^2}{\gamma^2}\left[Q-f_R|\psi|^2\right],\label{eq4}\ \ \\ &&
\xi(X)=\frac{1}{N}\sum_{m=-\infty}^{+\infty}\exp\left(-\frac{[X+mL]^2}{\sigma_X^2}\right),\label{eq5}
\label{eqom1}
\end{eqnarray}
where $\psi$, $\Delta_T$, $Q$ are the optical, thermal, and molecular vibrational fields, respectively. The above model assumes that only one radial mode family of the cylinder is at play, which may be unambiguously achieved by considering a hollow cylinder with an thin wall width, e.g., $w=0.75\ \mu$m, as we used  in Fig.\ref{f6}. The cylinder's dispersion is given by
\begin{eqnarray}
&& %\partial_t\psi(x,z,t)=-i\hat{D}\psi(x,z,t)\\ &&
\hat{D}\equiv\sum_{q=0}^{+\infty}\sum_{p=q}^{+\infty}b_{q,p-q}(-i\partial_z)^{p-q}(-i\partial_X)^{q},\\ &&
b_{q,p-q}\equiv\frac{B_{q,p-q}\gamma^{p/2-1}}{2^{p/2}|B_{0,2}|^{(p-q)/2}|B_{2,0}|^{q/2}},
\\ &&
B_{q,p-q}\equiv\frac{\tau\omega^{(q,p-q)}}{q!(p-q)!(2\pi R)^p},\\ &&
\omega^{(q,p-q)}\equiv\left.\partial_{k_x}^q\partial_{k_z}^{p-q}\omega(k_{x},k_{z})\right|_{k_{x0},k_{z0}},
\end{eqnarray}
where $\gamma$ is the normalised cavity loss, $R$ is the cylinder's radius, $\tau$ the roundtrip time, and $k_x$, $k_z$ the wavenumbers associated to the the $X$, $z$ coordinates ($X$ is the frame at rest in the lab). The dispersion terms with low $p,q$ indices account for: $p=0,q=0$, resonance frequency ($\omega_0$); $p=1,q=1$, group velocity of the pump's frequency along $X$ ($b_{1,1}$); $p=2,q=2$, GVD; $p=2,q=0$, diffraction. The rest of terms are all included in the \textit{higher order dispersion} operator $\hat{D}_{hod}\equiv\hat{D}-\omega_0\tau/\gamma+ib_{1,1}\partial_X-\frac{1}{2}(\partial_X^2-\partial_z^2)$. The even parity of the cylinder's dispersion around $z$ (see, e.g., \cite{Gorjosab,joannoPRL}) and the fact that we expand around the $k_{z0}=0$ yields the nullity of all coefficients with $p-q=1$. The relation between normalised and physical coordinates is as follows: $X=X_{phys}/(2\pi R)\sqrt{\gamma/(2|B_{2,0}|)}$, $z=Z_{phys}/(2\pi R)\sqrt{\gamma/(2|B_{0,2}|)}$.   
 
In our simulations, the width of the numerical window along $x$ was $L=16$, which together with the choice of $\gamma=0.001$ sets $B_{2,0}\approx-2\times10^{-6}$, attainable with a silica glass cylinder of $R=100\ \mu$m and wall-width $w=0.75\ \mu$m at $\lambda_p\approx1.26\ \mu$m, which features a roundtrip time $\tau=3.1$ ps and a quality factor $\mathcal{Q}=\omega_p\tau/\gamma\approx4.9\times10^6$ ($\omega_p=2\pi c/\lambda_p$), reasonable for cylinders \cite{sumcyl}. The main higher dispersion terms are given by $B_{3,0}\approx9.26\times10^{-10}$, $B_{4,0}\approx-2.83\times10^{-13}$, $B_{0,2}\approx1.21\times10^{-4}$. The normalised detuning is $\delta=(\omega_p-\omega_0)\tau/\gamma$.

The pump beam, assumed of Gaussian profile, has an amplitude $h_0$ and a width $\sigma_z$. Under realistic conditions (cf. Fig.\ref{f6}), the pump beam is also localised along $X$, and this is accounted for via the function $\xi(X)$, where $N$ is the normalisation factor such that $\max(\xi(X))=1$.

Thermal effects are introduced via the light-to-phonon energy conversion, $A=10^{-2}$, and the corresponding cooling rate, $B=5\times 10^{-2}$, considering realistic values \cite{menyuckther}. Thermal detuning, $\Delta_T$, is assumed not to depend on $X$ since $\tau\sim$ps is much smaller than the temperature diffusion time scale, in the order of the ns. Last, we note that in Eq.\ref{eq3} we omitted a term $\sim\mu\partial_z\Delta_T$, accounting for the heat diffusion along $z$, as its relative importance to the other terms is of the order of $\sim10^{-6}$ (the thermal diffusion coefficient is $\mu\approx7.25\times10^{-7}m^2/s$ \cite{kat}).

Raman scattering is introduced via the molecular vibrational field \cite{boyd}, previously implemented in microresonators \cite{milianPRA}, with standard parameter for glass given by \cite{blow}: Raman to Kerr effective fraction, $f_R=0.18$; inverse phonon life-time, $\gamma_R=1/32$ fs$^{-1}$; natural phonon frequency, $\omega_R=1/12.2$ fs$^{-1}$; and $\Omega_r\equiv[\gamma_R^2+\omega_R^2]^{1/2}$.

Eq.\ref{eq1} in the main text is a particular case of the system Eqns.\ref{eq2}-\ref{eq4} when higher order dispersion, Raman scattering, thermal detuning, and pump azimuthal's localisation are disregarded ($\hat{D}_{hod}=Q=\Delta_T=0$, $\xi(X)=1$). The time evolution of the above system, Eqns.\ref{eq2}-\ref{eq5}, is simulated via the fourth order Runge-Kutta method.

%The above model is rigorously obtained by applying the modal expansion approach \cite{chembo10,chembo13} to a cylindrical microresonator, yielding the corresponding generalisation of the Lugiato-Lefever model \cite{lugiato87}. An outline of the derivation may be found in \cite{IvarsPRL21} [see the corresponding supplemental material]. 

\ 

\noindent\textbf{Computation of stationary snakes.} 
Stationary solutions (snakes or else) are obtained numerically from Eq.\ref{eq1} or Eqns. \ref{eq2}-\ref{eq5} with the Newton-Raphson method in the frame comoving with the nonlinear state, where they readily satisfy $\partial_t\psi=0$. While equation \ref{eq1} is already expressed in such frame, Eqns.\ref{eq2}-\ref{eq5} (expressed in the lab frame) are rewritten into the comoving frame after the substitution $x=X-(b_{1,1}+v)t$, where $v$ is a velocity shift induced by the higher order effects, which is computed together with the nonlinear solution. When computing stationary states, the $x$-localisation of the pump is disregarded (we set $\xi(X)=1$).

%explain that a possibility is to superimpose the unstable mode with the background at some delta around the bifurcation point, so the obtained profile is used as an approximation for the quasi-stright snake (snake with low tilt $\alpha$). This approach is most useful to find bernaches were snakes are all unstable (like $S_1$ and $S_5$ in Fig.\ref{f2}a) so propagation simulations are of no help.

\ 

\noindent\textbf{Stability of snakes.} 
Linear stability analysis is performed to all stationary solutions, represented by the tuple $\{\psi_s\,Q_s,\Delta_{T_s}\}$. Each field in the stationary solution is prone to develop instabilities. In the initial stages of such instabilities fields are regarded as $\psi=\psi_s+ae^{i\lambda t}+b^*e^{-i\lambda^* t}$, $Q=Q_s+ce^{i\lambda t}+c^*e^{-i\lambda^* t}$, $\Delta_{T}=\Delta_{T_s}+de^{i\lambda t}+d^*e^{-i\lambda^* t}$ (Note $Q,\Delta_{T}$ are real fields), where $\lambda$ are the complex eigenvalues of the Jacobian matrix, obtained after substituion of the above decomposition into the system of Eqns. \ref{eq2}-\ref{eq4} (with $\xi(X)=1$) and linearising in $\lambda$. The real parts of $\lambda$, $\mathrm{Re}(\lambda)$, are the growth rates yielding instabilities when $\mathrm{Re}(\lambda)>0$.

\ 

\noindent\textbf{{Data availability}}

\noindent
The data that support the plots within this paper and other findings of this study are available from the corresponding author upon reasonable request.

\ 

\noindent\textbf{{Code availability}}

\noindent
The analysis codes will be made available on reasonable request.

\ 

\noindent\textbf{{Acknowledgements}}

\noindent
JAC and CM acknowledge support from the Spanish government via the Grant PID2021-124618NB-C21 funded by MCIN/AEI/ 10.13039/501100011033 and by “ERDF A way of making Europe”, by the “European Union”. CM acknowledges support from Generalitat Valenciana PROMETEO/2021/082. PFC acknowledges partial support from the Spanish gouvernement via the project PID2021-128676OB-I00 (MICINN). LT acknowledges support by CEX2019-000910-S [MCIN/AEI/10.13039/501100011033], Fundaci\'{o} Cellex, Fundaci\'{o} Mir Puig, and Generalitat de Catalunya (CERCA). YVK academic research has been supported by the research project FFUU-2021-0003 of the Institute of Spectroscopy of the Russian Academy of Sciences.

\ 

\noindent\textbf{{Author contributions}} 

\noindent
SBI and CM carried out the numerical simulations. CM conceived the project. All authors contributed significantly to this work, discussed the results, and contributed into the manuscript preparation.

\ 

\noindent\textbf{{Competing interests}} 

\noindent
The authors declare no competing interests.

%%% AN EXTRA BIBLIO SOURCE HERE?

%

\end{document}